%% file: paper.tex
\documentclass[a4paper]{article}
\usepackage[dvipdfmx]{graphicx}
\usepackage[cmex10]{amsmath}
\usepackage{amssymb}
\usepackage[scaled=0.92]{helvet}    % set Helvetica as the sans-serif font
\usepackage{cite}  
\usepackage{url}
\usepackage{bm}
\usepackage{color}
\usepackage{algpseudocode} 
\usepackage{tabularx,booktabs}

\newcolumntype{L}{>{\raggedright\arraybackslash}X}
\newcolumntype{C}{>{\centering\arraybackslash}X}
\newcolumntype{R}{>{\raggedleft\arraybackslash}X}

\input{mymacro} 

\title{Fast, Compact, and High Quality LSTM-RNN Based Statistical
       Parametric Speech Synthesizers for Mobile Devices}

\makeatletter
\def\name#1{\gdef\@name{#1\\}}
\makeatother \name{{\em Heiga~Zen,~Yannis~Agiomyrgiannakis,~Niels~Egberts,}
                   {\em Fergus~Henderson,~Przemys\l{}aw~Szczepaniak}}

\author{Heiga~Zen,~Yannis~Agiomyrgiannakis,~Niels~Egberts,\\
        Fergus~Henderson,~and~Przemys\l{}aw~Szczepaniak\\
        \\
        \GoogleLogo
        \\
        \texttt{\normalsize\{%
        heigazen,agios,nielse,fergus,pszczepaniak%
        \}@google.com
        }
}
\date{}
        
\begin{document}

\maketitle
\begin{abstract}
Acoustic models based on long short-term memory recurrent neural networks (LSTM-RNNs) were
applied to statistical parametric speech synthesis (SPSS) and showed significant improvements
in naturalness and latency over those based on hidden Markov models (HMMs).
This paper describes further optimizations of LSTM-RNN-based SPSS for deployment on mobile devices; 
weight quantization, multi-frame inference, and
robust inference using an $\epsilon$-contaminated Gaussian loss function.
Experimental results in subjective listening tests show that these optimizations can make
LSTM-RNN-based SPSS comparable to HMM-based SPSS in runtime speed while maintaining naturalness.
Evaluations between LSTM-RNN-based SPSS and HMM-driven unit selection speech synthesis are also presented.
\end{abstract}
\noindent{\bf Index Terms}: statistical parametric speech synthesis, recurrent neural networks.
\section{Introduction}
\label{sec:intro}
Statistical parametric speech synthesis (SPSS) \cite{Zen_SPSS_SPECOM}
based on artificial neural networks (ANN) has became popular in the text-to-speech (TTS) research area
in the last few years
\cite{
Zen_DNN_ICASSP,
Lu_DNN_SSW8,
Yao_DNN_ICASSP14,
Raitio_DNNsource_EUSIPCO2014,
Yin_DCT_Interspeech2014,
Zen_MDN_ICASSP14,
Fan_BLSTM_Interspeech14,
Zen_ULSTM_ICASSP,
Raul_BLSTM_Interspeech14,
Zhizhen_SimplifiedLSTM_ICASSP2016,
Hu_SinusoidalFusion_Interspeech2015,
Wu_MGEDNN_Interspeech2015,
Wu_DeepBN_ICASSP2015,
Watts_SentenceVector_Interspeech2015,
Fan_DNNAdapt_ICASSP,
Xie_MGE_Interspeech,
Yu_Investigation_ICSP,
Hashimoto_Effect_ICASSP,
Uria_TrajectoryRNADE_ICASSP2015
}.
ANN-based acoustic models offer an efficient and distributed representation of
complex dependencies between linguistic and acoustic features \cite{Zen_DeepLearn_SSW,Watts_DNN_ICASSP2016}
and have shown the potential to produce natural sounding synthesized speech
\cite{Zen_DNN_ICASSP,Yao_DNN_ICASSP14,Zen_MDN_ICASSP14,Fan_BLSTM_Interspeech14,Zen_ULSTM_ICASSP}.
Recurrent neural networks (RNNs) \cite{RNN}, especially long short-term memory (LSTM)-RNNs \cite{LSTM},
provide an elegant way to model speech-like sequential data that embodies
short- and long-term correlations.
They were successfully applied to acoustic modeling
for SPSS \cite{Fan_BLSTM_Interspeech14,Raul_BLSTM_Interspeech14,Zen_ULSTM_ICASSP,Zhizhen_SimplifiedLSTM_ICASSP2016}.
Zen \textit{et al.} proposed a streaming speech synthesis architecture using unidirectional
LSTM-RNNs with a recurrent output layer \cite{Zen_ULSTM_ICASSP}.
It enabled low-latency speech synthesis, which is essential in some applications.
However, it was significantly slower than hidden Markov model (HMM)-based SPSS \cite{yoshimura_PhD} 
in terms of real-time ratio \cite{Zen_ASRU2015}.
% which prevented full deployment of LSTM-RNN-based SPSS to a production environment.
This paper describes further optimizations of LSTM-RNN-based SPSS 
for deployment on mobile devices.
The optimizations conducted here include reducing computation and disk footprint, as well as
making it robust to errors in training data.
% Note that LSTM-RNN-based TTS with these optimizations was deployed in
% the Google TTS app for Android \cite{GoogleTTSApp} from version 3.8 (Feb. 2016).
% Google also deployed this configuration as main server-side TTS for several languages.

The rest of this paper is organized as follows.
Section~\ref{sec:lstm} describes the proposed optimizations. 
Experiments and subjective evaluation-based findings are presented in Section~\ref{sec:experiments}.
Concluding remarks are shown in the final section.

\begin{figure}[t]
  \centering
  \includegraphics[width=0.9\textwidth]{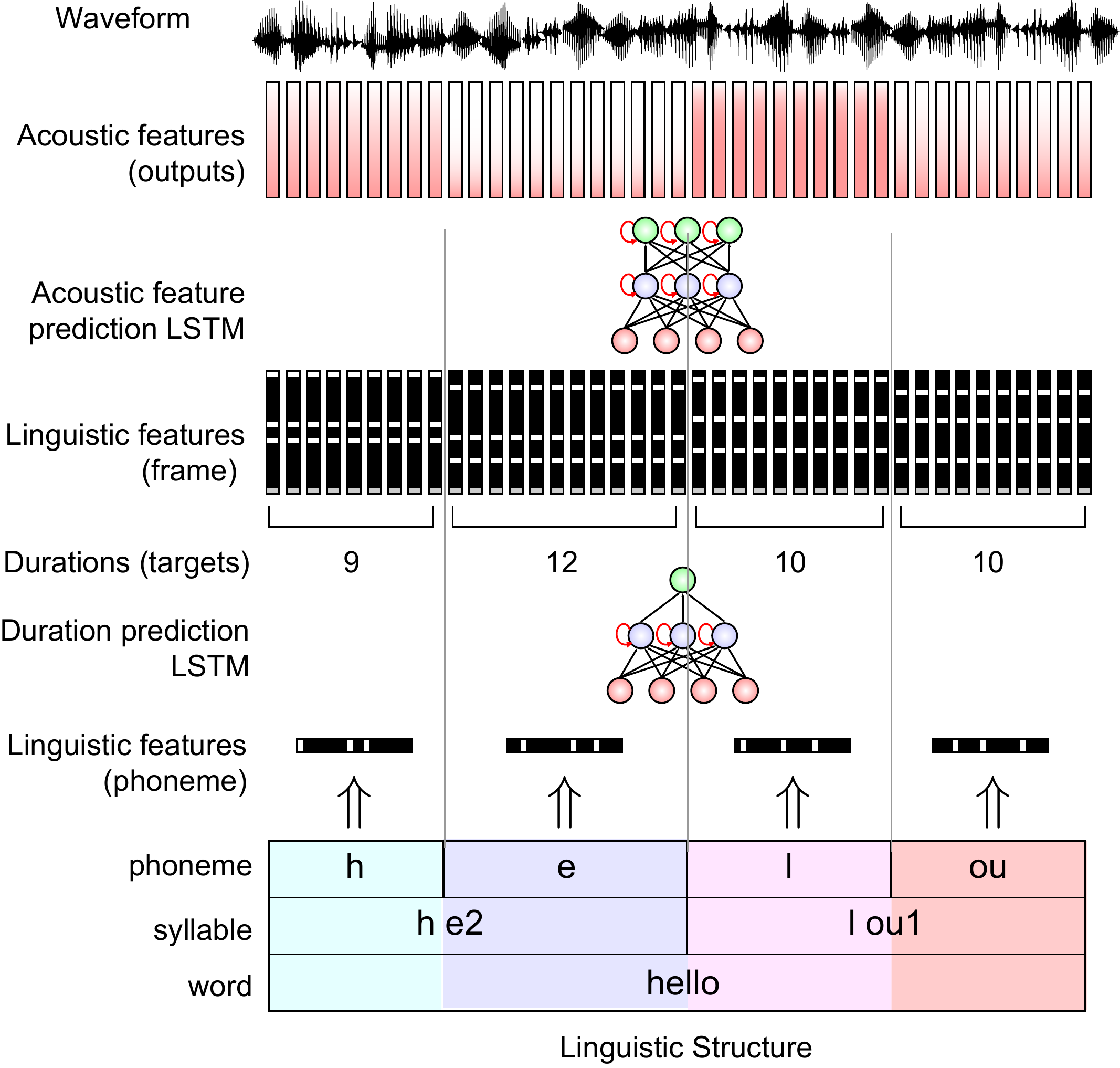}
  \caption{\label{fig:overview}
           \textit{
           Overview of the streaming SPSS architecture using LSTM-RNN-based acoustic and duration models \cite{Zen_ULSTM_ICASSP}.}}
\end{figure}%mm

\section{Optimizing LSTM-RNN-based SPSS}
\label{sec:lstm}

Figure~\ref{fig:overview} shows the overview of the streaming synthesis
architecture using unidirectional LSTM-RNNs \cite{Zen_ULSTM_ICASSP}.
Unlike HMM-based SPSS, which usually requires utterance-level batch processing \cite{tokuda_synHMM_ICASSP2000} or
frame lookahead \cite{Koishida_PARGEN_ICSP}, this architecture allows frame-synchronous streaming
synthesis with no frame lookahead.
Therefore this architecture provides much lower latency speech synthesis.
However, there are still a few drawbacks;
\begin{itemize}
  \item \textbf{Disk footprint};  
  Although the total number of parameters in LSTM-RNN-based SPSS can be significantly lower than that
  of HMM-based SPSS \cite{Zen_ULSTM_ICASSP}, the overall disk footprint of the LSTM-RNN system can be similar or slightly larger
  because HMM parameters can be quantized using 8-bit integers \cite{Phoneticarts_small_HMM}.
  Therefore decreasing the LSTM-RNN system disk footprint is essential for deployment
  on mobile devices.
  \item \textbf{Computation}; 
  With HMM-based SPSS, inference of acoustic parameters involves traversing decision trees at each
  {\em HMM state} and running the speech parameter generation algorithm \cite{tokuda_synHMM_ICASSP2000}.
  On the other hand, inference of acoustic parameters with LSTM-RNN-based SPSS
  involves many matrix-vector multiplications at each {\em frame}, which are expensive.
  This is particularly critical for client-side TTS on mobile devices,
  which have less powerful CPUs and limited battery capacity.
  \item \textbf{Robustness}; 
  Typical ANN-based SPSS relies on
  fixed phoneme- or state-level alignments \cite{Zen_DNN_ICASSP},
  whereas HMMs can be trained without fixed alignments using the Baum-Welch algorithm.
  Therefore, the ANN-based approach is less robust to alignment errors.
\end{itemize}
This section describes optimizations addressing these drawbacks.
Each of them that follow will be evaluated in Section~\ref{sec:experiments}.

\subsection{Weight quantization}
ANN weights are typically stored in 32-bit floating-point numbers.
However there are significant advantages in memory, disk footprint and
processing performance in representing them in lower integer precision.
This is commonly approached by quantizing the ANN weights.
This paper utilizes 8-bit quantization of ANN weights
\cite{Raziel_quantization_interspeech} to reduce the disk footprint of
LSTM-RNN-based acoustic and duration models.
Although it is possible to run inference in 8-bit integers with quantization-aware training
\cite{Raziel_quantization_interspeech}, that possibility is not utilized here;
instead weights are stored in 8-bit integer on disk then recovered to 32-bit
floating-point numbers after loading to memory.

\begin{figure}[t]
  \centering
  \includegraphics[width=0.9\textwidth]{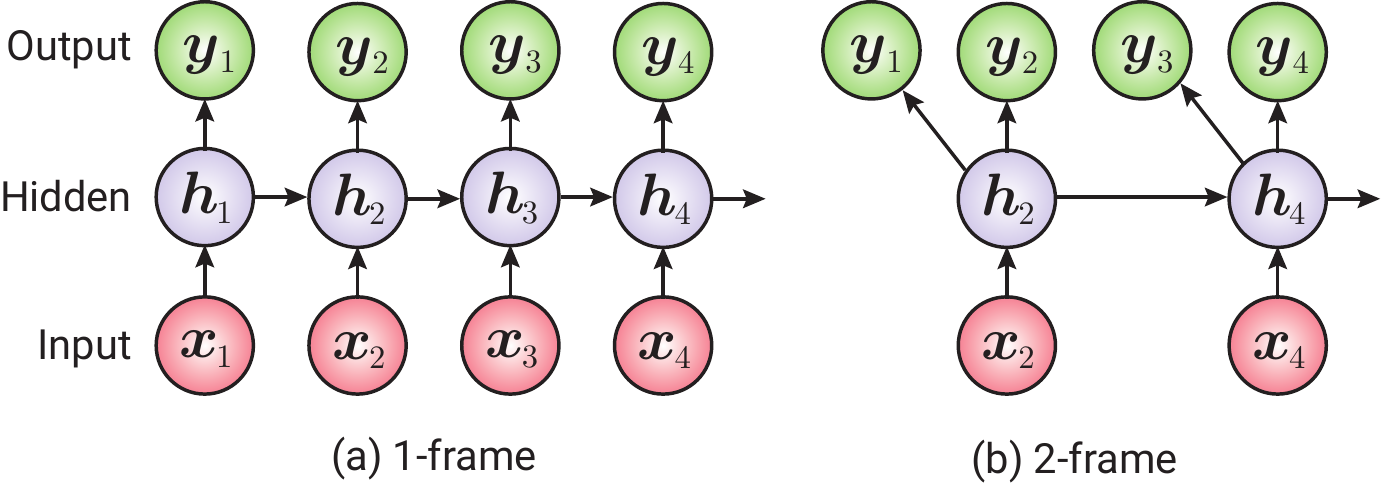}
  \caption{\label{fig:multiframe-lstm}
           \textit{
           Illustration of computation graph of (a) single-frame and (b) multi (two)-frame
  LSTM-RNNs.}}
\vspace{1mm}
  \includegraphics[width=0.9\textwidth]{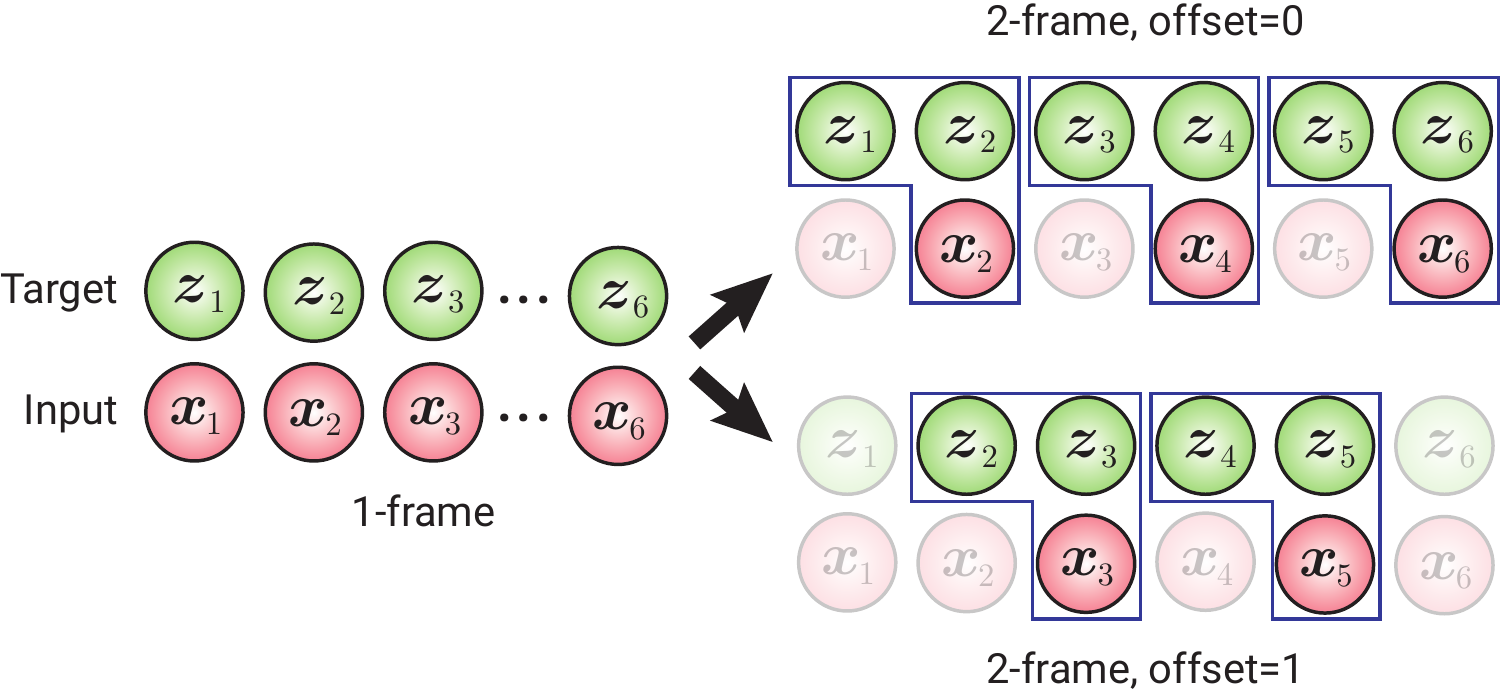}
  \caption{\label{fig:data_augmentation}
           \textit{
           Data augmentation with different offsets for 2-frame bundled inference.}}
\end{figure}%mm

\subsection{Multi-frame bundled inference} 
Inference of acoustic frames takes 60--70\% of total computations in our LSTM-RNN-based SPSS implementation.
Therefore, it is desirable to reduce the amount of computations at the inference stage.
In typical ANN-based SPSS, input linguistic features other than state- and frame-position features
are constant within a phoneme \cite{Zen_DNN_ICASSP}.
Furthermore, speech is a rather stationary process at 5-ms frame shift and target acoustic
frames change slowly across frames.
Based on these characteristics of inputs and targets this paper explores the multi-frame inference approach \cite{Vanhoucke_multiframe_ICASSP2013}.
Figure~\ref{fig:multiframe-lstm} illustrates the concept of multi-frame inference.
Instead of predicting one acoustic frame, multiple acoustic frames are jointly predicted at the same time instance.
This architecture allows significant reduction in computation while maintaining the
streaming capability.     

However, preliminary experiments showed degradation due to
mismatch between training and synthesis;
alignments between input/target features
can be different at the synthesis stage, e.g.,
training: $\bx_2 \to \{\by_1, \by_2\}$, synthesis: $\bx_3 \to \{\by_2, \by_3\}$.
This issue can be addressed by data augmentation.
Figure~\ref{fig:data_augmentation} shows the data augmentation with different frame offset.
From aligned input/target pairs, multiple data sequences can be generated with
different starting frame offset.
By using these data sequences for training, acoustic LSTM-RNNs will generalize
to different possible alignments between inputs and targets.

\begin{figure}[t]
  \centering
  \includegraphics[width=0.9\textwidth]{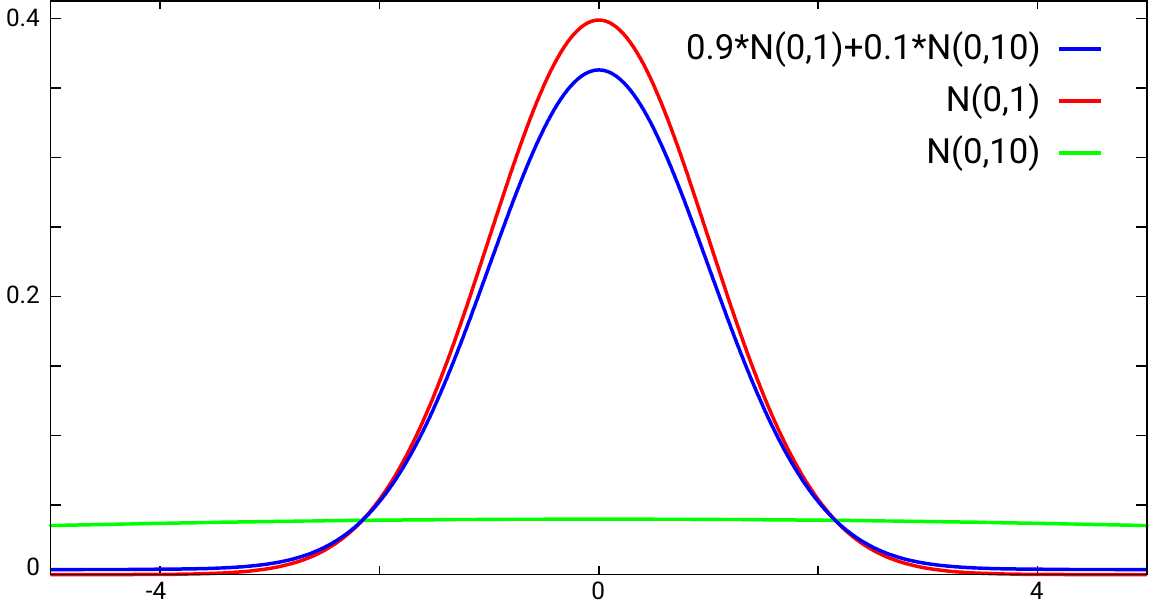} 
  \caption{\label{fig:contaminated}
           \textit{
           Plot of 1-dimensional $\epsilon$-contaminated Gaussian distribution
           ($\bmu=[0]$, $\bSigma=[1]$, $\epsilon=0.1$, $c=10$).}}
\end{figure}%mm

\subsection{Robust regression}
It is known that learning a linear regression model with the squared loss function can
suffer from the effect of outliers.
Although ANNs trained with the squared loss function are not a simple linear regression model,
their output layers perform linear regression given activations at the last hidden layer.
Therefore, ANNs trained with the squared loss function can be affected by outliers.
These outliers can come from recordings, transcriptions, forced alignments,
and $F_0$ extraction errors.

Using robust regression techniques such as
linear regression with a heavy-tailed distribution \cite{heavy_tail_loss_min} or
minimum density power divergence estimator \cite{Henter_RobustDNNDur_ICASSP}
can relax the effect of outliers.
In this work a simple robust regression technique
assuming that the errors follow a mixture of
two Gaussian distributions, in particular, $\epsilon$-contaminated Gaussian distribution \cite{contaminated_normal},
which is a special case of the Richter distribution \cite{Richter_Distribution,Brown_corrHMM_PhD,Gales_Richter_Eurospeech}, is employed;
the majority of observations are from a specified Gaussian distribution,
though a small proportion are from a Gaussian distribution with much higher variance,
while the two Gaussian distributions share the same mean.
The loss function can be defined as %defined as a negative log probability as
\begin{multline}
\mathcal{L}(\bz; \bx, \bLambda) = -
\log \bigl\{ (1-\epsilon) \Gauss \left(\bz ; f(\bx; \bLambda), \bSigma \right) % \\
+ \epsilon \Gauss \left(\bz ; f(\bx; \bLambda), c\bSigma \right) \bigr\},
\end{multline}
where $\bz$ and $\bx$ denote target and input vectors,
$\bSigma$ is a covariance matrix,
$\epsilon$ and $c$ are weight and scale of outliers, 
$\Lambda$ is a set of neural network weights, 
and $f(\cdot)$ is a non-linear function to predict an output vector given the input vector.
Typically, $\epsilon < 0.5$ and $c > 1$.
Note that if $\epsilon = 0$ and $\bSigma=\bI$, the $\epsilon$-contaminated
Gaussian loss function is equivalent to the squared loss function.
Figure~\ref{fig:contaminated} illustrates $\epsilon$-contaminated Gaussian distribution
($\bmu = [0]$, $\bSigma = [1]$, $c=10$ and $\epsilon=0.1$).
It can be seen from the figure that the $\epsilon$-contaminated Gaussian distribution
has heavier tail than the Gaussian distribution.
As outliers will be captured by the Gaussian distribution with wider variances,
the estimation of means is less affected by these outliers.
Here using the $\epsilon$-contaminated Gaussian loss function as a criterion to train LSTM-RNNs 
is investigated for both acoustic and duration LSTM-RNNs.
Note that the $\epsilon$-contaminated Gaussian distribution is similar to globally tied distribution (GTD)
in \cite{Yu_GTD_ICASSP}.

\section{Experiments}
\label{sec:experiments}

\subsection{Experimental conditions}
Speech data from a female professional speaker was used to train
speaker-dependent % acoustic and duration
unidirectional LSTM-RNNs for each language.
The configuration for speech analysis stage and data preparation process were the same as
those described in \cite{Zen_ULSTM_ICASSP} except the use of speech at
22.05~kHz sampling rather than 16~kHz and 7-band aperiodicities rather than 5-band ones.

Both the input and target features were normalized to be zero-mean unit-variance in advance.
The architecture of the acoustic LSTM-RNNs was 
1 $\times$ 128-unit ReLU \cite{ReLU} layer followed by
3 $\times$ 128-cell LSTMP layers \cite{Hasim_LSTM_Interspeech14} with 64 recurrent projection units
with a linear recurrent output layer  \cite{Zen_ULSTM_ICASSP}.
The duration LSTM-RNN used a single LSTM layer with 64 cells with feed-forward output layer with
linear activation.
To reduce the training time and impact of having many silence
frames, 80\% of silence frames were removed from the training data.
Durations of the beginning and ending silences were excluded from 
the training data for the duration LSTM-RNNs. The
weights of the LSTM-RNNs were initialized randomly.  Then
they were updated to minimize the mean squared error between the
target and predicted output features.
A distributed CPU implementation of mini-batch asynchronous stochastic 
gradient descent (ASGD)-based truncated back propagation through time (BPTT)
\cite{BPTT} algorithm was used \cite{Hasim_LSTM_Interspeech14}.
The learning rates for the acoustic and duration LSTM-RNNs were
$10^{-5}$ and $10^{-6}$, respectively.
The learning rates were exponentially decreased over time \cite{Senior_AdaDecay_ICASSP}.
Training was continued until the loss value over the development
set converged.
The model architecture and hyper-parameters were used across all languages.

At the synthesis stage, durations and acoustic features were predicted from linguistic
features using the trained networks.
Spectral enhancement based on post-filtering in the cepstral domain
\cite{yoshimura_PhD} was applied to improve the naturalness of
 the synthesized speech.
From the acoustic features, speech waveforms were synthesized
using the Vocaine vocoder \cite{Vocaine}.

To subjectively evaluate the performance of the systems, preference tests
were also conducted.
100 utterances not included in the training data
were used for evaluation.
Each pair was evaluated by at least eight native speakers of each language.
The subjects who did not use headphones were excluded from the experimental results.
After listening to each pair of samples,
the subjects were asked to choose their preferred one, or they
could choose ``no preference'' if they did not have any preference.
% Should we mention the 7-point scale used for A/B tests?
Note that stimuli that achieved a statistically significant preference ($p<0.01$) are presented in
bold characters in tables displaying experimental results in this section.

\subsection{Experimental results for optimizations}
\subsubsection{Weight quantization}
Table~\ref{tab:quantization} shows the preference test result comparing LSTM-RNNs with and without weight quantization.
It can be seen from the table that the effect of quantization was negligible.
The disk footprint of the acoustic LSTM-RNN for English (NA) was reduced from 1.05 MBytes to 272 KBytes.

\begin{table}[t]
\centering
\caption{\label{tab:quantization}
         \textit{
         Subjective preference scores (\%)
         between LSTM-RNNs with (\texttt{int8}) and without (\texttt{float})
         8-bit quantization.
         Note that ``English (GB)'', ``English (NA)'', and ``Spanish (ES)'' indicate British English, North American
         English, and European Spanish, respectively.
         }}
\vspace{1mm}
% \begin{tabularx}{0.42\textwidth}{c * {4}{C}}
\begin{tabularx}{0.8\textwidth}{c * {4}{C}}
\toprule
Language   & \texttt{int8} & \texttt{float} & No pref. \\ \midrule
English (GB) & 13.0 & 12.2 & 74.8 \\
English (NA) & 8.0 & 10.0 & 82.0 \\
French       & 4.7 & 3.8 & 91.5 \\
German       & 12.5 & 8.8 & 78.7 \\
Italian      & 12.0 & 9.8 & 78.2 \\
Spanish (ES) & 8.8 & 7.5 & 83.7 \\
\bottomrule
\end{tabularx}%
\vspace{1mm}
\caption{\label{tab:multiframe}
         \textit{
         Subjective preference scores (\%)
         between LSTM-RNNs using
         4-frame bundled inference with data augmentation (\texttt{4-frame}) and
         single-frame inference (\texttt{1-frame}).}}
\vspace{1mm}
% \begin{tabularx}{0.42\textwidth}{c * {4}{C}}
\begin{tabularx}{0.8\textwidth}{c * {4}{C}}
\toprule
Language   & \texttt{4-frame} & \texttt{1-frame} & No pref. \\ \midrule
English (GB) & 25.7 & 20.2 & 54.2 \\
English (NA) & 8.5  & 6.2  & 85.3 \\
French       & 18.8 & 18.6 & 62.6 \\
German       & 19.3 & 22.2 & 58.5 \\ 
Italian      & 13.5 & 14.4 & 72.1 \\
Spanish (ES) & 12.8 & 17.0 & 70.3 \\
\bottomrule
\end{tabularx}%
\vspace{1mm}
\caption{\label{tab:cn_loss}
         \textit{
         Subjective preference scores (\%)
         between LSTM-RNNs trained with
         the $\epsilon$-contaminated Gaussian (\texttt{CG}) and squared (\texttt{L2}) loss functions.}}
\vspace{1mm}
% \begin{tabularx}{0.42\textwidth}{c * {4}{C}}
\begin{tabularx}{0.8\textwidth}{c * {4}{C}}
\toprule
Language   & \texttt{CG} & \texttt{L2} & No pref. \\ \midrule
English (GB) & \textbf{27.4} & 18.1 & 54.5 \\
English (NA) & 7.6 & 6.8 & 85.6 \\ 
French     & \textbf{24.6} & 15.9 & 59.5 \\
German     & 17.1 & 20.8 & 62.1 \\
Italian    & \textbf{16.0} & 10.6 & 73.4 \\
Spanish (ES) & 16.0 & 13.4 & 70.6 \\
\bottomrule
\end{tabularx}%
\end{table}%

\subsubsection{Multi-frame inference}
While training multi-frame LSTM-RNNs,
the learning rate needed to be reduced (from $10^{-5}$ to $2.5 \times 10^{-6}$)
as mentioned in \cite{Vanhoucke_multiframe_ICASSP2013}.
Table~\ref{tab:multiframe} shows the preference test result comparing single and multi-frame inference.
Note that weights of the LSTM-RNNs were quantized to 8-bit integers.
It can be seen from the table that LSTM-RNN with multi-frame inference with data augmentation achieved the same naturalness
as that with single-frame one.
Compared with \texttt{1-frame}, \texttt{4-frame} achieved about 40\% reduction of walltime at runtime synthesis.

\subsubsection{$\epsilon$-contaminated Gaussian loss function}
Although $c$, $\epsilon$, and $\bSigma$ could be trained with the network weights,
they were fixed to $c=10$, $\epsilon=0.1$, and $\bSigma=\bI$ for both acoustic and duration LSTM-RNNs.%\footnote{These values chosen via preliminary experiments.}
Therefore, the numbers of parameters of the LSTM-RNNs trained with the squared and $\epsilon$-contaminated
Gaussian loss functions were identical.
For training LSTM-RNNs with the $\epsilon$-contaminated Gaussian loss function,
the learning rate could be increased (from $2.5 \times 10^{-6}$ to $5 \times 10^{-6}$ for acoustic LSTM-RNNs,
from $10^{-6}$ to $5 \times 10^{-6}$ for duration LSTM-RNNs). 
From a few preliminary experiments,
the $\epsilon$-contaminated Gaussian loss function with a 2-block
structure was selected; 1) mel-cepstrum and aperiodicities, 2)
$\log F_0$ and voiced/unvoiced binary flag.
This is similar to the multi-stream HMM structure \cite{HTK} used in HMM-based
speech synthesis \cite{yoshimura_PhD}.
Table~\ref{tab:cn_loss} shows the preference test result comparing the squared and $\epsilon$-contaminated normal loss function to train LSTM-RNNs.
Note that all weights of the LSTM-RNNs were quantized to 8-bit integers
and 4-frame bundled inference was used.
It can be seen from the table that LSTM-RNN trained with the $\epsilon$-contaminated normal loss function achieved the same or better naturalness
than those with the squared loss function.

\subsection{Comparison with HMM-based SPSS}
% The Google TTS app \cite{GoogleTTSApp}, which provides the TTS functionality for Android devices,
% has been using HMM-based SPSS.
% Although LSTM-RNN-based SPSS without the optimizations described in this paper
% was included since its version 3.6 (July 2015),
% it was disabled on most of the devices due to its high computational cost.
% As the optimizations described in this paper can reduce the computational cost of LSTM-RNNs,
% we aimed to replace HMM-based SPSS by LSTM-RNN-based one.

The next experiment compared HMM- and LSTM-RNN-based SPSS 
with the optimizations described in this paper.
Both HMM- and LSTM-RNN-based acoustic and duration models were quantized into 8-bit integers.
The same training data and text processing front-end modules were used.

The average disk footprints of HMMs and LSTM-RNNs including both acoustic and duration models
over 6 languages were 1560 and 454.5 KBytes, respectively. 
Table~\ref{tab:lstm-vs-hmm-latency} shows the average latency (time to get the first chunk of audio)
and average total synthesis time (time to get the entire audio) of the HMM and LSTM-RNN-based SPSS systems (North American English)
to synthesize a character, word, sentence, and paragraph on a Nexus 6 phone.
Note that the execution binary was compiled for modern ARM CPUs having the NEON advanced single instruction, multiple data (SIMD) instruction set \cite{ARM_NEON}. 
To reduce the latency of the HMM-based SPSS system, the recursive version of the speech
parameter generation algorithm \cite{Koishida_PARGEN_ICSP}
with 10-frame lookahead was used.
It can be seen from the table that the LSTM-RNN-based system could synthesize speech with
lower latency and total synthesis time than the HMM-based system.
However, it is worthy noting that the LSTM-RNN-based system was 15--22\% slower than the HMM-based system in terms of the total synthesis time  
on old devices having ARM CPUs without the NEON instruction set (latency was still lower).
Table~\ref{tab:lstm-vs-hmm-preference} shows the preference test result comparing
the LSTM-RNN- and HMM-based SPSS systems.
It shows that the LSTM-RNN-based system could synthesize more naturally sounding
synthesized speech than the HMM-based one.

\begin{table}[t]
\centering
\caption{\label{tab:lstm-vs-hmm-latency}
         \textit{
         Average latency and total time in milliseconds to synthesize a character, word, sentence, and paragraph
         by the LSTM-RNN- (\texttt{LSTM}) and HMM-based (\texttt{HMM}) SPSS systems.
         }}
\vspace{1mm}
% \begin{tabularx}{0.45\textwidth}{c * {5}{C}}
\begin{tabularx}{0.8\textwidth}{c * {5}{R}}
\toprule
 &
\multicolumn{2}{p{0.3\linewidth}}{\centering Latency (ms)} &
\multicolumn{2}{p{0.3\linewidth}}{\centering Total (ms)} \\ \cmidrule{2-5}
Length & \texttt{LSTM} & \texttt{HMM} & \texttt{LSTM} & \texttt{HMM} \\ \midrule
char.  & 12.5          & 19.5         & 49.8          & 49.6  \\
word   & 14.6          & 25.3         & 61.2          & 80.5  \\
sent.  & 31.4          & 55.4         & 257.3         & 286.2 \\
para.  & 64.1          & 117.7        & 2216.1       & 2400.8 \\
\bottomrule
\end{tabularx}%
\vspace{1mm}
\caption{\label{tab:lstm-vs-hmm-preference}
         \textit{
         Subjective preference scores (\%)
         between the LSTM-RNN- and HMM-based SPSS systems .}}
\vspace{1mm}
% \begin{tabularx}{0.42\textwidth}{c * {4}{C}}
\begin{tabularx}{0.8\textwidth}{c * {4}{C}}
\toprule
Language   & \texttt{LSTM} & \texttt{HMM} & No pref. \\ \midrule
English (GB) & 31.6 & 28.1 & 40.3 \\
English (NA) & \textbf{30.6} & 15.9 & 53.5 \\
French     & \textbf{68.6} & 8.4 & 23.0  \\
German     & \textbf{52.8} & 19.3 & 27.9  \\
Italian    & \textbf{84.8} & 2.9 & 12.3 \\
Spanish (ES) & \textbf{72.6} & 10.6 & 16.8 \\
\bottomrule
\end{tabularx}%
\end{table}

% Footprint HMM  LSTM
% French: 1.58MB  437.06KB
% German: 1.75MB   469.09
% Italian: 1.61MB  457.38KB
% Spanish: 1.45MB  438.62KB
% GB English: 1.28MB  467.93KB
% NA English: 1.69MB  459.72KB

\subsection{Comparison with concatenative TTS}
The last experiment evaluated the HMM-driven unit selection TTS
\cite{Xavi_Barracuda_interspeech}
and LSTM-RNN-based SPSS with the optimizations described in this paper except quantization.
Both TTS systems used the same training data and text processing front-end modules.
Note that additional linguistic features which were only available 
with the server-side text processing front-end modules were used in both systems.
The HMM-driven unit selection TTS systems were built from speech at 16 kHz sampling.
Although LSTM-RNNs were trained from speech at 22.05 kHz sampling,
speech at 16 kHz sampling was synthesized at runtime using a resampling functionality in Vocaine \cite{Vocaine}.
These LSTM-RNNs had the same network architecture as the one described in the previous section.
They were trained with the $\epsilon$-contaminated Gaussian loss function and utilized 4-frame bundled inference.
Table~\ref{tab:lstm-vs-hybrid} shows
the preference test result.
It can be seen from the table that the LSTM-RNN-based SPSS systems were
preferred to the HMM-driven unit selection TTS systems 
in 10 of 26 languages, while there was no significant preference between them in 3 languages.
Note that the LSTM-RNN-based SPSS systems were 3--10\% slower but 1,500--3,500 times smaller in disk footprint
than the hybrid ones.

\begin{table}[t]
\centering
\caption{\label{tab:lstm-vs-hybrid}
         \textit{
         Subjective preference scores (\%)
         between the LSTM-RNN-based SPSS and
         HMM-driven unit selection TTS (\texttt{Hybrid}) systems.
         Note that ``Spanish (NA)'' and ``Portuguese (BR)'' indicate
         North American Spanish and Brazilian Portuguese,
         respectively.
         }}
\vspace{1mm}
% \begin{tabularx}{0.42\textwidth}{c * {4}{C}}
\begin{tabularx}{0.8\textwidth}{c * {4}{C}}
\toprule
Language   & \texttt{LSTM} & \texttt{Hybrid} & No pref. \\ \midrule
Arabic     & 13.9 & \textbf{22.1} & 64.0 \\
Cantonese  & \textbf{25.1} & 7.3 & 67.6 \\
Danish     & 37.0 & \textbf{49.1} & 13.9 \\
Dutch      & 29.1 & \textbf{46.8} & 24.1 \\
English (GB) & 22.5 & \textbf{65.1} & 12.4 \\
English (NA) & 23.3 & \textbf{61.8} & 15.0 \\
French     & 28.4 & \textbf{50.3} & 21.4 \\
German     & 20.8 & \textbf{58.5} & 20.8 \\
Greek      & \textbf{42.5} & 21.4 & 36.1 \\
Hindi      & 42.5 & 36.4 & 21.1 \\
Hungarian  & \textbf{56.5} & 30.3 & 13.3 \\
Indonesian & 18.9 & \textbf{57.8} & 23.4 \\
Italian    & 28.1 & \textbf{49.0} & 22.9 \\
Japanese   & \textbf{47.4} & 28.8 & 23.9 \\
Korean     & \textbf{40.6} & 25.8 & 33.5 \\
Mandarin   & \textbf{48.6} & 17.5 & 33.9 \\
Norwegian  & \textbf{54.1} & 30.8 & 15.1 \\
Polish     & 14.6 & \textbf{75.3} & 10.1 \\
Portuguese (BR) & 31.4 & 37.8 & 30.9 \\
Russian    & 26.7 & \textbf{49.1} & 24.3 \\
Spanish (ES) & 21.0 & \textbf{47.1} & 31.9 \\
Spanish (NA) & 22.5 & \textbf{55.6} & 21.9 \\
Swedish    & \textbf{48.3} & 33.6 & 18.1 \\
Thai       & \textbf{71.3} & 8.8  & 20.0 \\
Turkish    & \textbf{61.3} & 20.8 & 18.0 \\
Vietnamese & 30.8 & 30.8 & 38.5 \\
\bottomrule
\end{tabularx}% 
\end{table}% 
 
\section{Conclusions}
\label{sec:conclusions}
This paper investigated three optimizations of LSTM-RNN-based SPSS
for deployment on mobile devices;
% \begin{enumerate}
1) Quantizing LSTM-RNN weights to 8-bit integers reduced disk footprint by 70\%, with no significant difference in naturalness;
2) Using multi-frame inference reduced CPU use by 40\%, again with no significant difference in naturalness;
3) For training, using an $\epsilon$-contaminated Gaussian loss function rather than a squared loss function to avoid excessive effects from outliers proved beneficial, allowing for an increased learning rate and improving naturalness.
% \end{enumerate}
The LSTM-RNN-based SPSS systems with these optimizations surpassed the HMM-based SPSS systems
in speed, latency, disk footprint, and naturalness on modern mobile devices.
Experimental results also showed that the LSTM-RNN-based SPSS system with the optimizations could match the HMM-driven unit selection 
TTS systems in naturalness in 13 of 26 languages.

\section{Acknowledgement}
The authors would like to thank Mr. Raziel Alvarez for helpful comments and discussions.

% \clearpage
\bibliographystyle{IEEEbib}
% \footnotesize
% \bibliography{references-short}

\end{document}

%% file: mymacro.tex
\bmdefine{\bO}{O}
\bmdefine{\bC}{C}
\bmdefine{\bc}{c}
\bmdefine{\bo}{o}
\bmdefine{\bW}{W}
\bmdefine{\bmu}{\mu}
\bmdefine{\bQ}{Q}
\bmdefine{\bq}{q}
\bmdefine{\bw}{w}
\bmdefine{\bU}{U}
\bmdefine{\bL}{L}
\bmdefine{\bu}{u}
\bmdefine{\bZero}{0}
\bmdefine{\bI}{I}
\bmdefine{\bR}{R}
\bmdefine{\bP}{P}
\bmdefine{\br}{r}
\bmdefine{\bmm}{m}
\bmdefine{\bsigma}{\sigma}
\bmdefine{\bSigma}{\Sigma}
\bmdefine{\bOmega}{\Omega}
\bmdefine{\bomega}{\omega}
\bmdefine{\bS}{S}
\bmdefine{\bA}{A}
\bmdefine{\bC}{C}
\bmdefine{\bM}{M}
\bmdefine{\bg}{g}
\bmdefine{\bs}{s}
\bmdefine{\bpsi}{\psi}
\bmdefine{\bPsi}{\Psi}
\bmdefine{\bphi}{\phi}
\bmdefine{\bPhi}{\Phi}
\bmdefine{\bPi}{\Pi}
\bmdefine{\bpi}{\pi}
\bmdefine{\bLambda}{\Lambda}
\bmdefine{\blambda}{\lambda}
\bmdefine{\bB}{B}
\bmdefine{\bb}{b}
\bmdefine{\bl}{l}
\bmdefine{\bd}{d}
\bmdefine{\bD}{D}
\bmdefine{\bY}{Y}
\bmdefine{\bG}{G}
\bmdefine{\bp}{p}
\bmdefine{\bxi}{\xi}
\bmdefine{\bmeta}{\eta}
\bmdefine{\bzeta}{\zeta}
\bmdefine{\bk}{k}
\bmdefine{\bK}{K}
\bmdefine{\bF}{F}
\bmdefine{\bv}{v}
\bmdefine{\bX}{X}
\bmdefine{\bx}{x}
\bmdefine{\by}{y}
\bmdefine{\bz}{z}
\bmdefine{\bZ}{Z}
\bmdefine{\bcalX}{\mathcal{X}}
\bmdefine{\bH}{H}
\bmdefine{\bh}{h}
\bmdefine{\bcalH}{\mathcal{H}}
\bmdefine{\bV}{V}

\def\Gauss{\mathcal{N}}

\definecolor {GoogleRed}   {rgb}{0.97265625, 0.00390625, 0.00390625}
\definecolor {GoogleBlue}  {rgb}{0.0078125,  0.3984375,  0.78125}
\definecolor {GoogleYellow}{rgb}{0.9453125,  0.70703125, 0.05859375}
\definecolor {GoogleGreen} {rgb}{0.0,        0.57421875, 0.23046875}
\def\GoogleLogo{\sf \textcolor{GoogleBlue}{G}\textcolor{GoogleRed}{o}\textcolor{GoogleYellow}{o}\textcolor{GoogleBlue}{g}\textcolor{GoogleGreen}{l}\textcolor{GoogleRed}{e}}

%% file: paper.bbl
\begin{thebibliography}{10}

\bibitem{Zen_SPSS_SPECOM}
H.~Zen, K.~Tokuda, and A.~Black,
\newblock ``Statistical parametric speech synthesis,''
\newblock {\em Speech Commn.}, vol. 51, no. 11, pp. 1039--1064, 2009.

\bibitem{Zen_DNN_ICASSP}
H.~Zen, A.~Senior, and M.~Schuster,
\newblock ``Statistical parametric speech synthesis using deep neural
  networks,''
\newblock in {\em Proc. ICASSP}, 2013, pp. 7962--7966.

\bibitem{Lu_DNN_SSW8}
H.~Lu, S.~King, and O.~Watts,
\newblock ``Combining a vector space representation of linguistic context with
  a deep neural network for text-to-speech synthesis,''
\newblock in {\em Proc. ISCA SSW8}, 2013, pp. 281--285.

\bibitem{Yao_DNN_ICASSP14}
Y.~Qian, Y.~Fan, W.~Hu, and F.~Soong,
\newblock ``On the training aspects of deep neural network ({DNN}) for
  parametric {TTS} synthesis,''
\newblock in {\em Proc. ICASSP}, 2014, pp. 3857--3861.

\bibitem{Raitio_DNNsource_EUSIPCO2014}
T.~Raitio, H.~Lu, J.~Kane, A.~Suni, M.~Vainio, S.~King, and P.~Alku,
\newblock ``Voice source modelling using deep neural networks for statistical
  parametric speech synthesis,''
\newblock in {\em Proc. EUSIPCO}, 2014, pp. 2290--2294.

\bibitem{Yin_DCT_Interspeech2014}
X.~Yin, M.~Lei, Y.~Qian, F.~Soong, L.~He, Z.-H. Ling, and L.-R. Dai,
\newblock ``Modeling {DCT} parameterized {F}0 trajectory at intonation phrase
  level with {DNN} or decision tree,''
\newblock in {\em Proc. Interspeech}, 2014, pp. 2273--2277.

\bibitem{Zen_MDN_ICASSP14}
H.~Zen and A.~Senior,
\newblock ``Deep mixture density networks for acoustic modeling in statistical
  parametric speech synthesis,''
\newblock in {\em Proc. ICASSP}, 2014, pp. 3872--3876.

\bibitem{Fan_BLSTM_Interspeech14}
Y.~Fan, Y.~Qian, and F.~Soong,
\newblock ``{TTS} synthesis with bidirectional {LSTM} based recurrent neural
  networks,''
\newblock in {\em Proc. Interspeech}, 2014, pp. 1964--1968.

\bibitem{Zen_ULSTM_ICASSP}
H.~Zen and H.~Sak,
\newblock ``Unidirectional long short-term memory recurrent neural network with
  recurrent output layer for low-latency speech synthesis,''
\newblock in {\em Proc. ICASSP}, 2015, pp. 4470--4474.

\bibitem{Raul_BLSTM_Interspeech14}
R.~Fernandez, A.~Rendel, B.~Ramabhadran, and R.~Hoory,
\newblock ``Prosody contour prediction with long short-term memory,
  bi-directional, deep recurrent neural networks,''
\newblock in {\em Proc. Interspeech}, 2014, pp. 2268--272.

\bibitem{Zhizhen_SimplifiedLSTM_ICASSP2016}
Z.~Wu and S.~King,
\newblock ``Investigating gated recurrent neural networks for speech
  synthesis,''
\newblock in {\em Proc. ICASSP}, 2016, pp. 5140--5144.

\bibitem{Hu_SinusoidalFusion_Interspeech2015}
Q.~Hu, Z.~Wu, K.~Richmond, J.~Yamagishi, Y.~Stylianou, and R.~Maia,
\newblock ``Fusion of multiple parameterisations for {DNN}-based sinusoidal
  speech synthesis with multi-task learning,''
\newblock in {\em Proc. Interspeech}, 2015, pp. 854--858.

\bibitem{Wu_MGEDNN_Interspeech2015}
Z.~Wu and S.~King,
\newblock ``Minimum trajectory error training for deep neural networks,
  combined with stacked bottleneck features,''
\newblock in {\em Proc. Interspeech}, 2015.

\bibitem{Wu_DeepBN_ICASSP2015}
Z.~Wu, C.~Valentini-Botinhao, O.~Watts, and S.~King,
\newblock ``Deep neural networks employing multi-task learning and stacked
  bottleneck features for speech synthesis,''
\newblock in {\em Proc. ICASSP}, 2015, pp. 4460--4464.

\bibitem{Watts_SentenceVector_Interspeech2015}
O.~Watts, Z.~Wu, and S.~King,
\newblock ``Sentence-level control vectors for deep neural network speech
  synthesis,''
\newblock in {\em Proc. Interspeech}, 2015, pp. 2217--2221.

\bibitem{Fan_DNNAdapt_ICASSP}
Y.~Fan, Y.~Qian, F.~Soong, and L.~He,
\newblock ``Multi-speaker modeling and speaker adaptation for {DNN}-based {TTS}
  synthesis,''
\newblock in {\em Proc. ICASSP}, 2015, pp. 4475--4479.

\bibitem{Xie_MGE_Interspeech}
F.-L. Xie, Y.~Qian, Y.~Fan, F.~Soong, and H.~Li,
\newblock ``Sequence error ({SE}) minimization training of neural network for
  voice conversion,''
\newblock in {\em Proc. Interspeech}, 2014, pp. 2283--2287.

\bibitem{Yu_Investigation_ICSP}
Z.~Chen and K.~Yu,
\newblock ``An investigation of implementation and performance analysis of
  {DNN} based speech synthesis system,''
\newblock in {\em Proc. ICSP}, 2014, pp. 577--582.

\bibitem{Hashimoto_Effect_ICASSP}
K.~Hashimoto, K.~Oura, Y.~Nankaku, and K.~Tokuda,
\newblock ``The effect of neural networks in statistical parametric speech
  synthesis,''
\newblock in {\em Proc. ICASSP}, 2015, pp. 4455--4459.

\bibitem{Uria_TrajectoryRNADE_ICASSP2015}
B.~Uria, I.~Murray, S.~Renals, and C.~Valentini-Botinhao,
\newblock ``Modelling acoustic feature dependencies with artificial neural
  networks: {T}rajectory-{RNADE},''
\newblock in {\em Proc. ICASSP}, 2015, pp. 4465--4469.

\bibitem{Zen_DeepLearn_SSW}
H.~Zen,
\newblock ``Deep learning in speech synthesis,''
  \url{http://research.google.com/pubs/archive/41539.pdf},
\newblock Invited keynote given at ISCA SSW8 2013.

\bibitem{Watts_DNN_ICASSP2016}
O.~Watts, G.~Henter, T.~Merritt, Z.~Wu, and S.~King,
\newblock ``From {HMM}s to {DNN}s: where do the improvements come from?,''
\newblock in {\em Proc. ICASSP}, 2016, pp. 5505--5509.

\bibitem{RNN}
A.~Robinson and F.~Fallside,
\newblock ``Static and dynamic error propagation networks with application to
  speech coding,''
\newblock in {\em Proc. NIPS}, 1988, pp. 632--641.

\bibitem{LSTM}
S.~Hochreiter and J.~Schmidhuber,
\newblock ``Long short-term memory,''
\newblock {\em Neural Comput.}, vol. 9, no. 8, pp. 1735--1780, 1997.

\bibitem{yoshimura_PhD}
T.~Yoshimura,
\newblock {\em Simultaneous modeling of phonetic and prosodic parameters, and
  characteristic conversion for {HMM}-based text-to-speech systems},
\newblock Ph.D. thesis, Nagoya Institute of Technology, 2002.

\bibitem{Zen_ASRU2015}
H.~Zen,
\newblock ``Acoustic modeling for speech synthesis: from {HMM} to {RNN},''
  \url{http://research.google.com/pubs/pub44630.html},
\newblock Invited talk given at ASRU 2015.

\bibitem{tokuda_synHMM_ICASSP2000}
K.~Tokuda, T.~Yoshimura, T.~Masuko, T.~Kobayashi, and T.~Kitamura,
\newblock ``Speech parameter generation algorithms for {HMM}-based speech
  synthesis,''
\newblock in {\em Proc. ICASSP}, 2000, pp. 1315--1318.

\bibitem{Koishida_PARGEN_ICSP}
K.~Koishida, K.~Tokuda, T.~Masuko, and T.~Kobayashi,
\newblock ``Vector quantization of speech spectral parameters using statistics
  of dynamic features,''
\newblock in {\em Proc. ICSP}, 1997, pp. 247--252.

\bibitem{Phoneticarts_small_HMM}
A.~Gutkin, J.~Gonzalvo, S.~Breuer, and P.~Taylor,
\newblock ``Quantized {HMM}s for low footprint text-to-speech synthesis,''
\newblock in {\em Proc. Interspeech}, 2010, pp. 837--840.

\bibitem{Raziel_quantization_interspeech}
R.~Alvarez, R.~Prabhavalkar, and A.~Bakhtin,
\newblock ``On the efficient representation and execution of deep acoustic
  models,''
\newblock in {\em Proc. Interspeech}, 2016.

\bibitem{Vanhoucke_multiframe_ICASSP2013}
V.~Vanhoucke, M.~Devin, and G.~Heigold,
\newblock ``Multiframe deep neural networks for acoustic modeling,''
\newblock in {\em Proc. ICASSP}, 2013, pp. 7582--7585.

\bibitem{heavy_tail_loss_min}
D.~Hsu and S.~Sabato,
\newblock ``Loss minimization and parameter estimation with heavy tails,''
\newblock {\em arXiv:1307.1827}, 2013.

\bibitem{Henter_RobustDNNDur_ICASSP}
G.~Henter, S.~Ronanki, O.~Watts, M.~Wester, Z.~Wu, and S.~King,
\newblock ``Robust {TTS} duration modeling using {DNN}s,''
\newblock in {\em Proc. ICASSP}, 2016, pp. 5130--5134.

\bibitem{contaminated_normal}
J.~Tukey,
\newblock ``A survey of sampling from contaminated distributions,''
\newblock {\em Contributions to probability and statistics}, vol. 2, pp.
  448--485, 1960.

\bibitem{Richter_Distribution}
A.~Richter,
\newblock ``Modelling of continuous speech observations,''
\newblock Tech. {R}ep. Advances in Speech Processing Conference, IBM Europe
  Institute, 1986.

\bibitem{Brown_corrHMM_PhD}
P.~Brown,
\newblock {\em The acoustic modeling problem in automatic speech recognition},
\newblock Ph.D. thesis, Carnegie Mellon University, 1987.

\bibitem{Gales_Richter_Eurospeech}
M.~Gales and P.~Olsen,
\newblock ``Tail distribution modelling using the {R}ichter and power
  exponential distributions,''
\newblock in {\em Proc. Eurospeech}, 1999, pp. 1507--1510.

\bibitem{Yu_GTD_ICASSP}
K.~Yu, T.~Toda, M.~G\u{a}s\'{i}c, S.~Keizer, F.~Mairesse, B.~Thomson, and
  S.~Young,
\newblock ``Probablistic modelling of {F}0 in unvoiced regions in {HMM} based
  speech synthesis,''
\newblock in {\em Proc. ICASSP}, 2009, pp. 3773--3776.

\bibitem{ReLU}
M.~Zeiler, M.~Ranzato, R.~Monga, M.~Mao, K.~Yang, Q.-V. Le, P.~Nguyen,
  A.~Senior, V.~Vanhoucke, J.~Dean, and G.~Hinton,
\newblock ``On rectified linear units for speech processing,''
\newblock in {\em Proc. ICASSP}, 2013, pp. 3517--3521.

\bibitem{Hasim_LSTM_Interspeech14}
H.~Sak, A.~Senior, and F.~Beaufays,
\newblock ``Long short-term memory recurrent neural network architectures for
  large scale acoustic modeling,''
\newblock in {\em Proc. Interspeech}, 2014, pp. 338--342.

\bibitem{BPTT}
P.~Werbos,
\newblock ``Backpropagation through time: what it does and how to do it,''
\newblock {\em Proc. IEEE}, vol. 78, no. 10, pp. 1550--1560, 1990.

\bibitem{Senior_AdaDecay_ICASSP}
A.~Senior, G.~Heigold, M.~Ranzato, and K.~Yang,
\newblock ``An empirical study of learning rates in deep neural networks for
  speech recognition,''
\newblock in {\em Proc. ICASSP}, 2013, pp. 6724--6728.

\bibitem{Vocaine}
Y.~Agiomyrgiannakis,
\newblock ``Vocaine the vocoder and applications is speech synthesis,''
\newblock in {\em Proc. ICASSP}, 2015, pp. 4230--4234.

\bibitem{HTK}
S.~Young, G.~Evermann, M.~Gales, T.~Hain, D.~Kershaw, X.-Y. Liu, G.~Moore,
  J.~Odell, D.~Ollason, D.~Povey, V.~Valtchev, and P.~Woodland,
\newblock ``The hidden {M}arkov model toolkit ({HTK}) version 3.4,''
  \url{http://htk.eng.cam.ac.uk/}, 2006.

\bibitem{ARM_NEON}
V.~Reddy,
\newblock ``{NEON} technology introduction,''
\newblock {\em ARM Corporation}, 2008.

\bibitem{Xavi_Barracuda_interspeech}
J.~Gonzalvo, S.~Tazari, C.-A. Chan, M.~Becker, A.~Gutkin, and H.~Silen,
\newblock ``Recent advances in {G}oogle real-time {HMM}-driven unit selection
  synthesizer,''
\newblock in {\em Proc. Interspeech}, 2016.

\end{thebibliography}
